\documentclass[10pt,journal,compsoc]{IEEEtran}
\usepackage{graphicx}
\usepackage{multirow}
\usepackage{dblfloatfix} 
\usepackage[dvipsnames]{xcolor}
\usepackage{soul, url}
\usepackage{color}
\usepackage{boldline}
\usepackage{xcolor,colortbl}

\usepackage{tikz}
\def\checkmark{\tikz\fill[scale=0.4](0,.35) -- (.25,0) -- (1,.7) -- (.25,.15) -- cycle;}
\usepackage{pifont}

\newcommand{\xmark}{\text{\ding{55}}}

\definecolor{Gray}{gray}{0.85}
\definecolor{LightCyan}{gray}{0.85}

\newcolumntype{g}{>{\columncolor{Gray}}c}

\usepackage[separate-uncertainty = true,multi-part-units = repeat]{siunitx}
\usepackage{array,amsfonts,amsmath}
\usepackage[font=small]{caption}
\usepackage{subcaption}
\ifCLASSOPTIONcompsoc
  \usepackage[nocompress]{cite}
\else
  \usepackage{cite}
\fi
\ifCLASSINFOpdf
\else
\fi
\begin{document}
\title{Multitask Learning from Augmented Auxiliary Data for Improving Speech Emotion Recognition}

\author{Siddique Latif, Rajib Rana, Sara Khalifa, Raja Jurdak~\IEEEmembership{Senior Member,~IEEE}, and \\Bj\"{o}rn W.\ Schuller,~\IEEEmembership{Fellow,~IEEE}
\IEEEcompsocitemizethanks{
\IEEEcompsocthanksitem  S.\ Latif is affiliated with USQ, Australia and Distributed Sensing Systems Group, Data61, CSIRO Australia. 
\IEEEcompsocthanksitem R. Rana is with University of Southern Queensland (USQ), Australia.
\IEEEcompsocthanksitem  S.\ Khalifa is affiliated with Distributed Sensing Systems Group, Data61, CSIRO Australia.
\IEEEcompsocthanksitem  R.\ Jurdak is with the Trusted Network Lab, and Chair in Applied Data Sciences at  Queensland University of Technology (QUT), Australia.
\IEEEcompsocthanksitem  B.\ Schuller is affiliated with GLAM -- the Group on Language, Audio, and Music, Imperial College London, UK, and the ZD.B Chair of Embedded Intelligence for Health Care and Wellbeing, University of Augsburg, Germany.

Corresponding E-mail: siddique.latif@usq.edu.au}}

\IEEEtitleabstractindextext{%
\begin{abstract}
Despite the recent progress in speech emotion recognition (SER), state-of-the-art systems lack generalisation across different conditions. A key underlying reason for poor generalisation is the scarcity of emotion datasets, which is a significant roadblock to designing robust machine learning (ML) models. Recent works in SER focus on utilising multitask learning (MTL) methods to improve generalisation by learning shared representations. However, most of these studies propose MTL solutions with the requirement of meta labels for auxiliary tasks, which limits the training of SER systems. This paper proposes an MTL framework (MTL-AUG) that learns generalised representations from augmented data. We utilise augmentation-type classification and unsupervised reconstruction as auxiliary tasks, which allow training SER systems on augmented data without requiring any meta labels for auxiliary tasks. The semi-supervised nature of MTL-AUG allows for the exploitation of the abundant unlabelled data to further boost the performance of SER. We comprehensively evaluate the proposed framework in the following settings: (1) within corpus, (2) cross-corpus and cross-language, (3) noisy speech, (4) and adversarial attacks. Our evaluations using the widely used IEMOCAP, MSP-IMPROV, and EMODB datasets show improved results compared to existing state-of-the-art methods.

\end{abstract}

}

\maketitle

\IEEEdisplaynontitleabstractindextext
\IEEEpeerreviewmaketitle

\section{Introduction
\label{sec:introduction}}

\IEEEPARstart{S}{peech} Emotion Recognition (SER) is an emerging area of research. Speech contains information about human emotions, which can be utilised by machine learning (ML) systems for automatic detection redefining human-computer interactions. SER can help improve the quality of customer service by tracking customer-agent reactions. In healthcare, SER can be used for diagnosis and monitoring of affective behaviours \cite{latif2020speech,rana2019automated}. Service delivery in transport \cite{zepf2020driver}, forensics \cite{tavi2020prosodic}, education \cite{yadegaridehkordi2019affective}, media \cite{vanderplaetse2020improved} can be improved by utilising SER. 

Human emotion modelling is quite complex due to its dependency on many factors including speaker \cite{aldeneh2021you}, gender \cite{nediyanchath2020multi}, age \cite{wang2017learning}, culture \cite{latif2018cross}, and dialect \cite{laukka2014evidence}. Researchers have explored various ML techniques, including hidden Markov models, support vector machines, and deep neural networks (DNNs) for SER, wherein DNNs have improved performance compared to the classical ML techniques. Deep belief networks (DBN) \cite{hinton2006fast}, convolutional neural networks (CNN) \cite{lecun1989handwritten}, and  recurrent neural network (RNNs) have been successful in modelling emotions in speech and widely explored in SER \cite{latif2018variational,latif2020deep,neumann2017attentive,latif2021survey}. In particular, 
RNN architectures like short term memory (LSTM) networks \cite{hochreiter1997long} or bidirectional LSTM (BLSTM) combined with CNNs are a popular choice in SER for capturing emotional attributes and have been explored by many researchers \cite{atila2021attention,latif2019direct}. Studies \cite{latif2020deep,tzirakis2018end} show that the CNN-LSTM can learn better emotional features for SER compared to using CNN or LSTM individually. This work presents a unique semi-supervised configuration using CNN-BLSTM with attention mechanisms. We utilise an attention mechanism in our emotion classifier to combine the important emotional information extracted from the overall utterance and improve emotion classification performance.


Literature shows, SER models lack generalisation due to the single task-specific training and perform poorly when the data mismatch increases between the training and testing phases \cite{latif2020multi,zhang2019attention}. Typically, generalisation of deep learning models 
is improved by training them on diverse data. For example, state-of-the-art models in computer vision are trained on thousands of labelled samples, and automatic speech recognition systems are trained on 
thousands of hours of transcribed data \cite{latif2020deeprepre,malik2021automatic,latif2022survey}. In contrast, SER corpora are relatively small, and the creation of emotional corpora is a time consuming and expensive task \cite{parthasarathy2020semi,latif2020multi}
as emotion is subjective, and several annotators are usually required, which often have to repeatedly go through the speech material to annotate, e.\,g.,  affective dimension by affective dimension. 
To obtain data volume, most existing studies in SER attempt to train models on multiple corpora \cite{latif2018cross,latif2018transfer}. However, standard benchmark datasets are also very limited, which creates tremendous barriers to achieving generalisation in SER systems \cite{latif2021survey}.

An alternative technique to improve the generalisation of DL models is multitask learning (MTL) \cite{caruana1997multitask}, which simultaneously solves the multiple relevant auxiliary tasks along with the primary task. MTL can use different aspects of the same data or get data supplement from the secondary tasks. In this way, models can be better regularised by capturing shared and essential high-level representations, leading to an improved generalisation of the system. MTL has been successfully used in SER by achieving promising performance. However, most of these MTL techniques present \textit{supervised} auxiliary tasks, which require accurate annotations just like the primary emotion recognition tasks. Examples include emotional attributes (i.e., arousal, valence, and dominance) prediction \cite{xia2017multi,parthasarathy2017jointly}, gender identification  \cite{kim2017multi,latif2020multi,li2019improved}, speaker recognition \cite{latif2020multi,peri2021disentanglement}, and secondary emotion learning \cite{Lotfian2018}. The MTL methods with any of the above auxiliary tasks need accurate meta labels that limit the SER models' training. In some scenarios, larger data can be utilised for auxiliary tasks like speaker and gender identification \cite{latif2020multi}; however, collecting speaker and gender labels is also time- and labour-intensive. This also makes the model's performance speaker-dependent in some cases. Moreover, 
a 
generalised representation for SER containing speaker and gender information might be used maliciously without the user's consent by an eavesdropping adversary  \cite{ali2021privacy,jaiswal2020privacy}.  

In this paper, we propose a semi-supervised MTL framework that learns from augmented data---we call it MTL-AUG. It primarily classifies emotions and utilises \textit{data augmentation-type classification} and \textit{unsupervised reconstruction} as auxiliary tasks to learn generalised representations. We use types of augmentation as labels for 
\textit{data augmentation for classification} as an auxiliary task. 
In this way, these auxiliary tasks do not require meta labelling performed by experts. Our idea is inspired by ConvNets, which learn image classification features by predicting the 2D image rotation that is applied to the input image \cite{gidaris2018unsupervised}. Such geometric transformation cannot be applied to the speech signal. Therefore, we propose to use speech-based augmentation 
types 
that enable multitask training to learn a generalised representation without requiring meta labels. We apply temporal, frequency, and mixup related augmentations to the input speech. This allows the model to learn temporal and frequency related variations applied to the input data through augmentation-type classification as an auxiliary task. Learning the temporal and frequency variations in the data helps the MTL model 
to improve SER performance. Our second auxiliary task of unsupervised reconstruction acts as a regulariser and improves the quality of learnt representations. Overall, both auxiliary tasks 
enable the proposed framework to effectively utilise the augmented and unlabelled data to improve the generalisation of the SER system.

Most of previous MTL studies \cite{parthasarathy2017jointly,xia2017multi,tao2018advanced,li2019improved,latif2020multi,parthasarathy2020semi,peri2021disentanglement} evaluate the proposed models in within-corpus SER, and very few studies perform cross-corpus and cross-language SER. Moreover, none of these studies performs evaluations in noisy and adversarial attack settings. This is mainly due to the complexity of mismatch conditions in noisy and adversarial attacks. To show the advantage of our proposed MTL framework, we rigorously evaluate it against noisy and adversarial conditions. For evaluation, we use three widely used emotional databases: The interactive emotional dyadic motion capture (IEMOCAP) \cite{busso2008iemocap} database, MSP-IMPROV \cite{busso2017msp}, and the EMODB data. We compare our framework's performance with multiple recent studies and baseline CNN-BLSTM implementations. The comparative results in within-corpus, cross-corpus, cross-language, noisy and adversarial settings show that the proposed MTL-AUG framework achieves considerably improved performance, which attests to the strong generalisation power of the proposed MTL-AUG framework.

\section{Related Work}

\subsection{Multi-task Learning for SER}
Multitask learning (MTL) \cite{caruana1997multitask} aims to improve the generalisation of models by learning the similarities and differences among the given tasks from the training data. It has been successful to produce shared representation by simultaneously modelling multiple related tasks. The conventional single task learning technique ignores the information of related tasks and can increase the risk of overfitting \cite{zhang2019attention}. In contrast, MTL acts as a regulariser to reduce the risk of overfitting by introducing an inductive bias. Several MTL approaches \cite{li2014heterogeneous,zhang2014facial,farabet2013learning} have been exploited in computer vision to address various problems with significantly improved results. The speech community also explored MTL approaches to improve the performance of the tasks, including automatic speech recognition \cite{ravanelli2020multi}, speaker identification \cite{chen2015multi}, and also emotion classification \cite{xia2015multi}.




\begin{table*}[!ht]
\centering
\tiny
 \caption{Summary of a comparative analysis of our paper with that of the existing literature.}
\begin{tabular}{|l|l|ll|lllll|}
\hline
  
  & & \multicolumn{2}{c|}{\textbf{Label independent auxiliary tasks}}        
  & \multicolumn{5}{c|}{Evaluations}\\ \cline{3-9}
\multirow{-2}{*}{Paper/Author (Year)} &  \multirow{-2}{*}{\textbf{Label dependent auxiliary tasks}} &\multicolumn{1}{l|}{\begin{tabular}[c]{@{}l@{}}Reconstruction\end{tabular}}& \cellcolor[HTML]{C0C0C0}\begin{tabular}[c]{@{}l@{}}Augmentation-type \\ classification\end{tabular} &
\multicolumn{1}{l|}{\begin{tabular}[c]{@{}l@{}}within-\\ corpus\end{tabular}} &
\multicolumn{1}{l|}{\begin{tabular}[c]{@{}l@{}}Cross-\\ corpus\end{tabular}} &
\multicolumn{1}{c|}{\begin{tabular}[c]{@{}c@{}}Cross- \\ language\end{tabular}} &
\multicolumn{1}{c|}{\cellcolor[HTML]{C0C0C0}\begin{tabular}[c]{@{}c@{}}Noisy \\ conditions\end{tabular}} & \multicolumn{1}{c|}{\cellcolor[HTML]{C0C0C0}\begin{tabular}[c]{@{}c@{}}Adversarial \\ attacks\end{tabular}} \\ \hline

\begin{tabular}[l]{@{}l@{}}Prthasarathy and \\ Busso \cite{parthasarathy2017jointly} (2017)   \end{tabular} & \multicolumn{1}{l|}{ \begin{tabular}[c]{@{}l@{}} emotional attributes prediction    \end{tabular}}  & \multicolumn{1}{l|}{\xmark{}}      & \cellcolor[HTML]{C0C0C0} \xmark{}  & \multicolumn{1}{l|}{\checkmark{}}  &  \multicolumn{1}{l|}{\checkmark{}}    & \multicolumn{1}{l|}{\xmark{}}&\multicolumn{1}{l|}{\cellcolor[HTML]{C0C0C0}\xmark{}} & \cellcolor[HTML]{C0C0C0}   \xmark{}                                                                                 \\ \hline

Xia et al.\ \cite{xia2017multi} (2017)      & \multicolumn{1}{l|}{\begin{tabular}[c]{@{}l@{}} emotional attributes prediction    \end{tabular}}   & \multicolumn{1}{l|}{\xmark{}}       & \cellcolor[HTML]{C0C0C0} \xmark{}  & \multicolumn{1}{l|}{\checkmark{}}  & \multicolumn{1}{l|}{\checkmark{}}  &\multicolumn{1}{l|}{\xmark{}} &\multicolumn{1}{l|}{\cellcolor[HTML]{C0C0C0}\xmark{}}& \cellcolor[HTML]{C0C0C0}   \xmark{}   \\ \hline

Kim et al.\ \cite{kim2017towards} (2017)      & \multicolumn{1}{l|}{\begin{tabular}[c]{@{}l@{}} emotional attributes prediction +\\gender identification   \end{tabular}}           & \multicolumn{1}{l|}{\xmark{}}       & \cellcolor[HTML]{C0C0C0} \xmark{}  & \multicolumn{1}{l|}{\checkmark{}}  & \multicolumn{1}{l|}{\checkmark{}}  &\multicolumn{1}{l|}{\checkmark{}} &\multicolumn{1}{l|}{\cellcolor[HTML]{C0C0C0}\xmark{}}& \cellcolor[HTML]{C0C0C0}   \xmark{}   \\ \hline

Lotfian et al.\ \cite{Lotfian2018} (2018)   & \multicolumn{1}{l|}{\begin{tabular}[c]{@{}l@{}} emotional attributes classification    \end{tabular}}  & \multicolumn{1}{l|}{\xmark{}}  & \cellcolor[HTML]{C0C0C0} \xmark{}&  \multicolumn{1}{l|}{\checkmark{}}&   \multicolumn{1}{l|}{\xmark{}} &\multicolumn{1}{l|}{\xmark{}} &\multicolumn{1}{l|}{\cellcolor[HTML]{C0C0C0}\xmark{}}& \cellcolor[HTML]{C0C0C0}   \xmark{}                                                                                 \\ \hline
Tao et al.\ \cite{tao2018advanced} (2018)       & \multicolumn{1}{l|}{\begin{tabular}[c]{@{}l@{}} speaker classification +\\gender classification   \end{tabular}}         & \multicolumn{1}{l|}{\xmark{}}       & \cellcolor[HTML]{C0C0C0} \xmark{}   & \multicolumn{1}{l|}{\checkmark{}} & \multicolumn{1}{l|}{\xmark{}}    & \multicolumn{1}{l|}{\xmark{}} &\multicolumn{1}{l|}{\cellcolor[HTML]{C0C0C0}\xmark{}}& \cellcolor[HTML]{C0C0C0}   \xmark{}                                                                                 \\ \hline
Li et al. \cite{li2019improved} (2019)    & \multicolumn{1}{l|}{\begin{tabular}[c]{@{}l@{}}gender identification   \end{tabular}}          & \multicolumn{1}{l|}{\xmark{}}      & \cellcolor[HTML]{C0C0C0} \xmark{}    & \multicolumn{1}{l|}{\checkmark{}}& \multicolumn{1}{l|}{\xmark{}}    & \multicolumn{1}{l|}{\xmark{}} & \multicolumn{1}{l|}{\cellcolor[HTML]{C0C0C0}\xmark{}}&\cellcolor[HTML]{C0C0C0}   \xmark{}                                                                                 \\ \hline
\begin{tabular}[l]{@{}l@{}}Prthasarathy and \\ Busso \cite{parthasarathy2020semi} (2020)\end{tabular}          & \multicolumn{1}{l|}{\begin{tabular}[c]{@{}l@{}} emotional attributes prediction    \end{tabular}}        & \multicolumn{1}{l|}{\checkmark{}}      & \cellcolor[HTML]{C0C0C0} \xmark{}  & \multicolumn{1}{l|}{\checkmark{}}  & \multicolumn{1}{l|}{\checkmark{}}    & \multicolumn{1}{l|}{\xmark{}} &\multicolumn{1}{l|}{\cellcolor[HTML]{C0C0C0}\xmark{}}& \cellcolor[HTML]{C0C0C0}   \xmark{}                                                                                 \\ \hline
Latif et al.\ \cite{latif2020multi} (2020)        & \multicolumn{1}{l|}{\begin{tabular}[c]{@{}l@{}} speaker classification +\\gender classification   \end{tabular}}         & \multicolumn{1}{l|}{\checkmark{}}        & \cellcolor[HTML]{C0C0C0} \xmark{}  & \multicolumn{1}{l|}{\checkmark{}}  & \multicolumn{1}{l|}{\checkmark{}}    & \multicolumn{1}{l|}{\xmark{}} &\multicolumn{1}{l|}{\cellcolor[HTML]{C0C0C0}\xmark{}}& \cellcolor[HTML]{C0C0C0}   \xmark{}                                                                                 \\ \hline
Peri et al.\ \cite{peri2021disentanglement} (2021)     & \multicolumn{1}{l|}{\begin{tabular}[c]{@{}l@{}} speaker identification  \end{tabular}}   & \multicolumn{1}{l|}{\xmark{}}       & \cellcolor[HTML]{C0C0C0} \xmark{} & \multicolumn{1}{l|}{\checkmark{}}   & \multicolumn{1}{l|}{\xmark{}}    & \multicolumn{1}{l|}{\xmark{}} &\multicolumn{1}{l|}{\cellcolor[HTML]{C0C0C0}\xmark{}}& \cellcolor[HTML]{C0C0C0}   \xmark{}                                                                                 \\ \hline
\rowcolor[HTML]{C0C0C0} 
Our Paper (2022)                      & \multicolumn{1}{l|}{\textbf{None}\cellcolor[HTML]{C0C0C0}}  & \multicolumn{1}{l|}{\checkmark{}}  & \checkmark{}    & \checkmark{}& \multicolumn{1}{l|}{\checkmark{}\cellcolor[HTML]{C0C0C0}}  & \multicolumn{1}{l|}{\checkmark{}\cellcolor[HTML]{C0C0C0}}             &\checkmark{} &  \checkmark{}                                                                                                   \\ \hline
\end{tabular}
\label{table:litr}
\end{table*}

Eyben et al.\ \cite{eyben2012multitask} were the first to explore MTL in SER. They empirically found that multi-task training of models help improve performance in contrast to single-task training. Xia et al.\ \cite{xia2017multi} presented a DBN based MTL model for SER and utilised activation and valence labels as an auxiliary task. They demonstrated that the performance of SER for categorical emotion could be enhanced using activation and valence label information as auxiliary tasks. Prthasarathy and Busso \cite{parthasarathy2017jointly} presented a DNN-based MTL model that jointly learns the arousal, dominance, and valence from a given utterance. The authors found that joint training of the model with multiple emotional attributes enhances the performance compared to training with single attribute information. Ma et al.\ \cite{ma2018speech} used a multitask attention-based DNN model for SER and showed that a high performance could be achieved by optimising the model for joint classification of categorical emotions along with valence and activation labels classification. Similarly, Lotfian et al.\ \cite{Lotfian2018} utilised a DNN based framework for modelling primary and secondary emotions. Based on the results,  the authors showed that the performance of the primary classification task (categorical emotions) is enhanced by utilising the information of secondary emotions and emotional classes perceived by the evaluators. 

Another way to implement MTL in SER is to use speaker and gender identification as auxiliary tasks. Multiple studies have explored this phenomenon to improve SER performance. In \cite{tao2018advanced}, the authors presented an LSTM-based MTL framework that uses speaker and gender classification as auxiliary tasks to improve the performance of the main task, emotion classification. In another study \cite{latif2020multi}, the authors proposed an MTL framework that uses speaker and gender recognition as auxiliary tasks and used other speech corpora with speaker and gender labels and injected this data into the model. They showed that the performance could be significantly improved. Kim et al.\ \cite{kim2017towards} utilised gender and naturalness (natural or acted corpus) recognition as auxiliary tasks and evaluated the model using different corpora. They found that a performance gain can be achieved using gender or naturalness classification as auxiliary tasks. Other recent studies also utilised \cite{nediyanchath2020multi,li2019improved} gender-aware MTL SER models and found that emotion classification can be improved with additional gender label information.

Previous studies on MTL demonstrate that the use of auxiliary tasks helps improve SER performance compared with STL. However, these approaches either use information about emotional attributes (activation, valence, etc.) or non-emotional attributes (speaker, gender, etc.) that are not widely available in real-life. Also, labelling speech data with such meta-information is a cumbersome and expensive process. Some studies \cite{parthasarathy2020semi,latif2020multi} exploit the unsupervised reconstruction as auxiliary tasks; however, they also require additional labels for emotional attributes \cite{parthasarathy2020semi}, and gender and speaker labels in \cite{latif2020multi} for their MTL frameworks.

In contrast to previous studies, we propose an MTL framework that improves the performance without requiring such meta labels by annotators. We propose using data transformation (or augmentation)-type recognition and unsupervised feature reconstruction as auxiliary tasks. This allows us to utilise the type of augmentation applied to the input data as labels for the auxiliary task to train the proposed MTL framework.

\subsection{Data Augmentation in SER}
Data augmentation techniques have been used to generate additional training data for SER. For example, studies \cite{latif2019direct,aldeneh2017using} show that the speed perturbation \cite{ko2015audio} data augmentation technique can improve the performance of an SER system by generating copies of each utterance with different speed effects. The mixup \cite{zhang2018mixup} technique augments an SER system by generating the synthetic sample as a linear combination of the original sample. In SER, Latif et al. \cite{latif2020deep} augment the SER system with mixup to achieve robustness against noisy conditions. They showed that augmentation techniques make the training data diverse and help improve performance. A new method of data augmentation is SpecAugment \cite{park2019specaugment} and was proposed for automatic speech recognition, which is directly applied to the feature inputs of a neural network. In \cite{baird2021emotion}, the authors utilised the SpecAugment technique to augment their SER system with the duplicate samples by a factor of two. The authors highlighted that the data augmentation improves the robustness of the model by providing diverse training samples. Other studies \cite{latif2019direct,aldeneh2017using,latif2020federated} also achieve improved performance by exploiting data augmentation techniques to increase the training data. However, these studies only utilised the data augmentation in single-task learning to increase the training samples. In this paper, we propose to use data augmentation-type recognition as our auxiliary task in our proposed multitask learning framework. We hypothesise that multitask learning models are able to understand the concept of emotions while recognising the transformation performed on the input signal.

\subsection{SER robust to Adversarial Attacks and Noise}
In SER, it is essential to achieve robustness against perturbation/noise added to the input samples. However, very few studies focus on evaluating SER systems' robustness against noisy conditions and adversarial attacks. Huang et al.\  \cite{huang2017deep} used a CNN-LSTM model for robust SER. They found that CNN demonstrates a certain degree of noise robustness. In \cite{triantafyllopoulos2019towards}, the authors utilised deep residual networks for speech enhancement to remove noise from speech while preserving emotions for SER. Some other studies \cite{pandharipande2019front,juszkiewicz2014improving} also explored different noise removal frameworks for SER in noisy environments instead of achieving robustness by learning generalised representation. Based on the findings of data augmentation techniques to improve robustness \cite{zhang2017mixup,pang2019mixup}, a recent study \cite{latif2020deep} evaluated the regularising effect of data augmentation to improve the robustness of SER. They show that data augmentation helps to improve the robustness of SER against noise and adversarial attacks. However, no study has evaluated data augmentation in MTL scenarios to learn generalised representation to improve robustness in SER. 

\begin{figure*}[!ht]
\centering
\includegraphics[width=0.7\textwidth]{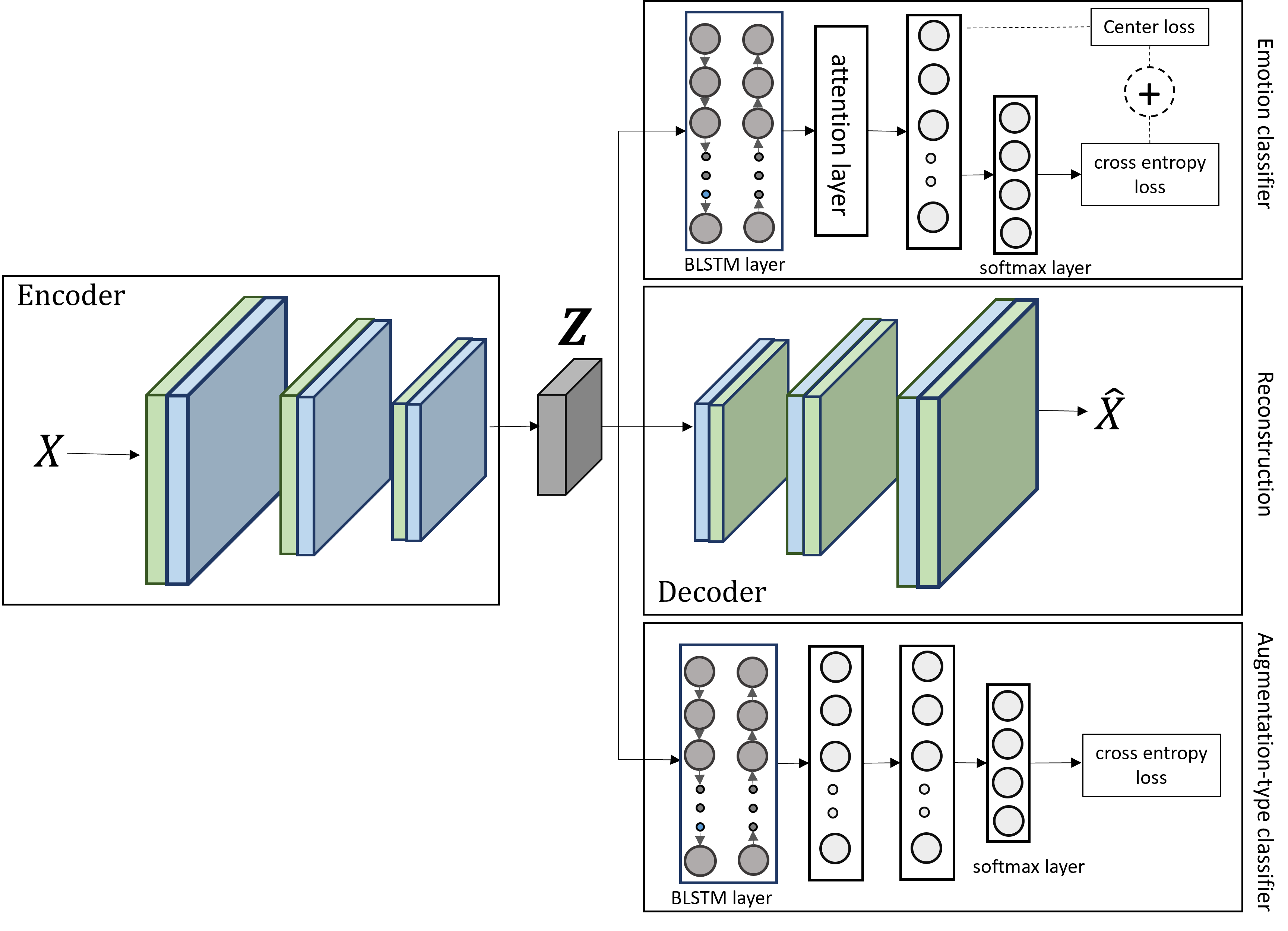}
\caption{Illustration of our proposed multitask framework for SER, which uses augmentation-type classification and reconstruction as auxiliary task to achieve better performance on the primary emotion classification task.}
\label{fig:Model}
\end{figure*}

\subsection{Summary}
We summarise the differences between our work and the existing literature in Table \ref{table:litr}. 
\begin{enumerate}
\item While some studies used reconstruction as an auxiliary task, no studies used augmentation-type classification as the auxiliary task. 
\item None of the studies evaluated their models' generalisation ability against noisy conditions and adversarial attacks. 
\item Most of the studies evaluated their model within-corpus settings by using training and testing data from the same corpus. Only a few studies evaluated the generalisation of proposed models in cross-corpus and even less in cross-language settings.

\end{enumerate}

\section{Methodology}
The proposed MTL-AUG framework uses the \textit{augmentation-type classification} and \textit{unsupervised reconstruction} as auxiliary tasks to learn generalised emotional representations. Before we describe our framework, we briefly introduce speech data augmentation, especially the techniques used for this work.
\subsection{Speech Data Augmentation}
We use augmentation to introduce variability and volume in the data. Speech signals can be augmented/transformed using different techniques. We use the following three techniques: (1) speed perturbation \cite{ko2015audio}, (2) mixup \cite{zhang2018mixup}, and (3) SpecAugment \cite{park2019specaugment}. \newline
\textbf{Speed perturbation} is a very popular and widely used audio augmentation technique that produces a warped time signal. Given a speech signal $x(t)$, time warping is performed by a factor $\alpha$ to produce the signal $x(\alpha t)$. In this way, speed perturbation changes the duration of a given speech signal. It can be applied directly on raw speech as we use in this paper.   \newline 
\textbf{SpecAugment} is used as a simple data augmentation method for Automatic Speech Recognition (ASR). 
It acts on the log-Mel spectrogram directly with a negligible amount of additional computational cost \cite{park2019specaugment}. In SpecAugment, training data can be augmented using spectro-temporal modifications to the original spectrograms by applying frequency and time masks. In frequency masking, a mask of size $f$ is chosen from a uniform distribution ($0$ to $F$) and consecutive log-Mel frequency channels $[f_{0}$, $f_{0}$+$f)$ are masked, where ${f_0}$ is chosen from $[0$, $v$ $-$ $f)$ and $v$ represents the number of Mel-frequency channels. In the time masking, a mask size of $t$ is chosen from a uniform distribution from $0$ to $T$, and the consecutive time steps $[t_{0}, t_{0}+t)$ are masked in time -- here, $t_{0}$ is chosen from $[0, \tau-t)$ and $\tau$ represents log-Mel spectrogram time steps. \newline
\textbf{Mixup} generates an augmented sample and its label by randomly mixing two inputs and their corresponding labels. This regularises the neural network to favour simple linear behaviour in-between training samples. It constructs augmented training examples as follows:
\begin{equation}
\label{m1}
    \tilde{x}=\lambda x_{i}+(1-\lambda) x_{j}
\end{equation}
\begin{equation}
\label{m2}
    \tilde{y}=\lambda y_{i}+(1-\lambda) y_{j},
\end{equation}
where ($x_{i}$, $y_{i}$) and ($x_{j}$, $y_{j}$) are randomly selected two examples from training data, and $\lambda$ $\in$ [0, 1].  Mixup can be applied on the features as well as on the raw speech \cite{zhang2018mixup}. We use mixup on Mel-spectrograms.

Data augmented using the above three techniques are fed to the proposed MTL-AUG framework to learn temporal, frequency, and mixup related changes applied to the data through the augmentation-type classification as auxiliary task. Note that, in SER, it is always important to capture spectro-temporal dynamics to accurately identify speech emotions \cite{kim2017learning,latif2020augmenting,latif2020deep}. In our proposed framework, we model spectro-temporal and augmentation related dynamics through auxiliary tasks in an MTL setting, which helps improve the performance of the primary emotion classification task. 
We will explain our proposed MTL-AUG framework next.

\subsection{MTL-AUG Framework}
Figure \ref{fig:Model} describes the proposed semi-supervised MTL architecture. Overall, the framework has four subnetworks: (1) encoder $E$, (2) decoder $D$, (3) emotion classifier $C_E$, and (4) augmentation-type classifier $C_A$. The proposed model is trained with MTL loss:  
\begin{equation}
  \mathcal{L}_{\text{MT}} = \mathcal{L}_{\text{pri}}+\lambda_{1}\mathcal{L}_{\text{aux}},
\end{equation}
where $\mathcal{L}_{\text{pri}}$ and $\mathcal{L}_{\text{aux}}$ represent the primary and auxiliary tasks, respectively. $\lambda_{1}$ is a hyper-parameter trading off primary and auxiliary tasks.

Our primary task is optimised with an emotion classifier $C_E$ that takes the encoded representation ($Z$) by the encoder ($E$) network to perform an emotion classification. It uses BLSTM layers for contextual modelling and an attention layer to combine the most salient features given to a dense layer for discriminative feature representation before classification. For a given output sequence $h_{i}$, utterance level important features are computed by the attention layer using:
\begin{equation}
    R_{\text{attentive}}=\sum_{i}\alpha_{i}h_{i},
\end{equation}
where $\alpha_{i}$ represents the attention weights that can be computed as follows:
\begin{equation}
    \alpha_{i}=\frac{\text{exp}W^T h_{i}}{\sum_{j}\text{exp}W^T h_{j}},
\end{equation}
where $W$ is a trainable parameter. The output attentive representation $R_{\text{attentive}}$ computed by the attention layer is fed to the dense layer for emotion classification. Our intuition of using the attention layer for SER is that the emotional content is distributed over the speech utterances. The attention layer weighs information extracted from different pieces of utterance and combines them into a weighted sum that helps produce better emotion classification performance \cite{neumann2017attentive}. The emotion classifier ($C_E$) is optimised using the sum of cross-entropy and centre loss functions: 
\begin{equation}
  \mathcal{L}_{\text{pri}} = \mathcal{L}_{\text{S}}+\lambda_{2}\mathcal{L}_{\text{C}},
\end{equation}
where $\mathcal{L}_{\text{S}}$ and $\mathcal{L}_{\text{C}}$ represent softmax cross-entropy loss and centre loss, respectively. $\lambda_{2}$ is the trade-off parameter between these two losses. The use of centre loss helps to minimise intra-class variations while maintaining separation between features of different classes by pulling them closer to their correspondence centres. The centre loss function can be defined as:
\begin{equation}
    \mathcal{L}_{\text{C}}=\frac{1}{1}\sum_{i=1}^{m}\lVert{f({x_i})-c_{y_i}}\rVert_{2}^{2}, 
\end{equation}
where $f({x_i})$ represents the deep features extracted from the last hidden layer and $c_{y_i}$ $\in$ $\mathbb{R}$ denotes $y_{i}^{th}$ class centre of the deep features. 

The secondary tasks in our framework are augmentation-type classification and reconstruction of the input speech features. In the reconstruction auxiliary task, the encoder and decoder networks minimise the reconstruction loss. 
The objective function for the autoencoder is:
\begin{equation}
\label{AE}
    \mathcal{L_{\text{AE}}}(x,D_{\theta}(E_{\theta}(x)))=\lVert{X-\hat{X}}\rVert_{2}^{2}.
\end{equation}

The other auxiliary task is to classify the transformation applied to the input. For this, we use classifier $C_A$ that takes the encoder $E$ output ($Z=E_{\theta}(x)$)  and performs  classification. 
We created augmented data by applying speed perturbation on raw speech, and the SpecAugment and mixup techniques to the Mel-spectrogram of emotional data samples. In augmentation-type  classification, we also consider samples with no augmentation as one class. Thus, classifier $C_A$ is trained on the four-way classification task of recognising one of the four classes (i.\,e., speech perturbation, SpecAugment, mixup, and no augmentation). 

The proposed framework is trained in a semi-supervised way as it uses both unsupervised and supervised learning \cite{deng2017semisupervised}. For the input $X$, the encoder network creates the latent code, which is an unsupervised process. The latent code is then used by the classifiers ($C_A$,$C_E$) with labels conforming to supervised learning.  
Note here that when using additional auxiliary data with no labels for emotion, the loss functions for augmentation-type classification
and the autoencoding network are only calculated to update the encoder network.


\section{Experimental Setup}

\subsection{Datasets}
\label{sec:data}
To evaluate the performance of our MTL-AUG model, we use three different datasets: IEMOCAP, EMODB, and MSP-IMPROV, which are commonly used for speech emotion classification research \cite{Lotfian+2016,kim2016emotion}. Both, the IEMOCAP and the MSP-IMPROV datasets are collected by simulating naturalistic dyadic interactions among professional actors and have similar labelling schemes. EMODB contains audio samples in the German language, and we use it for cross-language evaluations.  In order to use additional data for auxiliary tasks, we use the Librispeech~\cite{panayotov2015librispeech} dataset.

\subsubsection{IEMOCAP}
This is a multimodal database containing 12 hours of recorded data \cite{busso2008iemocap}. The recordings were collected during dyadic interactions from $10$ professional actors (five males and five females). Dyadic interactions allowed the actors to perform spontaneous emotion in contrast to reading text with prototypical emotions \cite{lotfian2017building}. Each interaction is around five minutes long and segmented into smaller utterances of sentences. Each sentence is annotated by the participant and three annotators for categorical labels.  Finally, an utterance is assigned a label if at least three annotators assigned the same label. Overall, this corpus contains nine emotions including angry, disgust, fearful, frustrated, sad, happy, excited, surprised, and neutral. Similar to prior studies \cite{latif2018variational,latif2019direct,latif2020multi}, we use utterances of four categorical emotions, including angry, happy, neutral, and sad in this study by merging ``happy'' and ``excited'' as one emotion class ``happy''. The final dataset includes $5531$ utterances ($1103$ angry,  $1708$ neutral, $1084$ sad, and $1636$ happy).     

\subsubsection{MSP-IMPROV}
This corpus is a multimodal emotional database recorded from $12$ actors performing dyadic interactions \cite{busso2017msp}, similar to IEMOCAP \cite{busso2008iemocap}. The utterances in MSP-IMPROV are grouped into six sessions, and each session has recordings of one male, and one female actor. The scenarios were carefully designed to promote naturalness while maintaining control over lexical and emotional contents. The emotional labels were collected through perceptual evaluations using crowdsourcing \cite{burmania2016increasing}. The utterances in this corpus are annotated in four categorical emotions: angry, happy, neutral, and sad. To be consistent with previous studies \cite{latif2019direct,gideon2017progressive},  we use all utterances with four emotions: anger (792), sad (885), neutral (3477), and happy (2644). 

\subsubsection{EMODB}
EMODB \cite{burkhardt2005database} is a popular and most widely used publicly available emotional dataset in the German Language. This corpus was recorded by the Institute of Communication Science, Technical University Berlin. EMODB contains audio recordings of seven emotions recorded by ten professional speakers in 10 German sentences. In this work, we select four basic emotions: angry, sad, neutral, and happy, to perform categorical cross-language SER as executed in \cite{latif2022self}. 

\subsubsection{LibriSpeech}
The LibriSpeech dataset \cite{panayotov2015librispeech} contains $1\,000$ hours of English read speech from $2\,484$ speakers. This corpus is derived from audiobooks and is commonly used for automatic speaker and speech recognition tasks \cite{berard2018end,dubey2019transfer}. The training portion of LibriSpeech is divided into three subsets, with an approximate recording time of $100$, $360$, and $500$ hours. Here, we choose the subset that contains $100$ hours of recordings and use it as additional unlabelled data. These recordings span over $251$ speakers. 

\subsubsection{DEMAND}
We select the Diverse Environments Multichannel Acoustic Noise Database (DEMAND) dataset \cite{thiemann2013diverse} as a source
of our noise signal. DEMAND contains audio recordings of various real-world noises recorded in various indoor and outdoor settings. In our experiments, we select noise recordings with 16\,kHz sampling rate to match with that of the audio recording of the speech emotion datasets.

\subsection{Features and Augmentation-Types}
We represent the speech utterances in log-Mel spectrograms, which is a popular 2D feature representation widely used for speech-related tasks, including SER. We apply overlapping Hamming windows with a size of 40\,ms and with a 10\,ms window shift. The height of the log-Mel spectrogram is 128. We set the length of utterances to 7.5\,s. Longer utterances are cut at 7.5\,s, and smaller utterances are padded with zeros. We select the length of the utterances based on validation results and previous studies \cite{li2019improved,latif2019direct}.

As outlined above, we apply three augmentation-types, including speed perturbation, mixup, and SpecAugment. For the speed perturbation, we create two copies of each training utterance by applying the speed effect at $0.9$ and $1.1$. We apply speed perturbation on the raw speech using the Sox\footnote{http://sox.sourceforge.net} audio manipulation tool, while we apply mixup and SpecAugment on the Mel spectrogram. 

\subsection{Hyperparameters}
For all the experiments, we use the Adam optimiser with default parameters. We start training models with a learning rate of $0.0001$ and calculate the validation accuracy at the end of each epoch. If the validation accuracy does not improve after five consecutive epochs, we halve the learning rate and restore the model to the best epoch. This process continues until the learning rate reaches below $0.00001$. We apply a rectified linear unit (ReLU) as a non-linear activation function type, as it gave us better performance than leaky ReLU and hyperbolic tangent during validation. 

Our {\bf baseline model} consists of the convolutional encoder network and Bidirectional LSTM (BLSTM)-based classification network. CNN layers in the encoder network produce the high-level feature representations. We use a larger kernel size for the first convolutional layer and reduce the kernel size in the remaining layers, as suggested by previous studies \cite{dai2019learning,gideon2019improving}. Feature representations learnt by the encoder network are given to the BLSTM layer with 128 LSTM units for emotional context modelling. After the BLSTM layer, we apply an attention layer to aggregate the emotional content distributed over the different parts of the given utterance. The attentive features are fed to the fully connected layer with 128 hidden units to produce emotionally discriminative features for a softmax layer. The softmax layer uses the crossentropy loss to produce the posterior class probabilities by enabling the network to learn separable features. In addition, we also exploit the centre loss to reduce the features' intra-class variation to improve the classification performance. 

In contrast to the baseline model, our MTL-AUG model contains two additional components: the decoder and augmentation-type classifier. The decoder network is used to reconstruct the input log-Mel spectrograms back from the encoded output by the encoder network. It has a similar architecture to the encoder, replacing convolutional layers with the transposed convolutional layers. The augmentation-type classifier takes the encoded representation and uses a BLSTM based classifier to classify different augmentation-types. We use one BLSTM layer with 256 LSTM units and two fully connected layers with 128 hidden units for auxiliary task classification. In addition, we use a dropout layer with a dropout rate of $0.3$ between two dense layers. We decide on the dropout rate based on validation experiments.

\section{Experiments and Results}

All the experiments are performed in a speaker-independent manner. In particular, we follow a easily reproducible leave-one-speaker-out cross-validation scheme commonly used in the literature \cite{latif2018variational,latif2020multi}. For cross-language SER, we follow \cite{latif2022self,kim2017towards} and use IEMOCAP and EMODB for a four-class emotion classification task. We use LibriSpeech as additional unlabelled data; results are presented in this section as ``MTL-AUG (additional data)''. For all the experiments,  we repeated each experiment ten times and calculated the mean and standard deviation. Results are presented using the unweighted average recall rate (UAR), a widely accepted metric in the field.

\subsection{Within Corpus Experiments} 
\label{within}
For the within-corpus setting, we compare the performance of the proposed model with the baseline. We also extend our evaluation by comparing the results with different multi-task learning approaches \cite{xia2015multi,neumann2019improving,latif2020multi} in Table \ref{table:within}. Our proposed MTL-AUG achieves better results than the baseline CNN-BLSTM architecture and other MTL approaches. Some studies \cite{xia2015multi,neumann2019improving} use dimensional emotion prediction as a secondary task to improve the classification of categorical emotions. They use additional information labels annotated by experts for dimensional emotions to perform an auxiliary task in their MTL frameworks. In another MTL study, \cite{latif2020multi}, speaker and gender identification are used as secondary tasks for shared generalised representation learning with multitasking semi-supervised adversarial autoencoder (SS-AAE). The authors also exploit the additional unlabelled data for the auxiliary task to boost the primary emotion classification task. However, this model also requires additional labels for speaker and gender and cannot exploit unlabelled data without this meta information. In contrast, we can utilise any speech data in the system without requiring information about the speaker and gender. In Table \ref{table:within}, we present these results with MTL-AUG (additional data) that performs augmentation-type classification and reconstruction as the auxiliary tasks on the additional speech from LibriSpeech to learn generalised representations. 
As our proposed auxiliary tasks do not require additional annotation by experts, it makes the MTL training more practical, yet better performing than the existing studies.

\begin{table}[!ht]
\centering
\caption{Comparison of results (UAR \%) of our proposed MTL-AUG framework with those of recent MTL studies. MTL-AUG
(additional data) represents when additional unlabelled data from LibriSpeech is used.}
\begin{tabular}{|l|c|c|}
\hline
Model & IEMOCAP & MSP\_IMPROV \\ \hline
DBN (MTL) \cite{xia2015multi} & 62.2        &  -           \\ \hline 
Attentive CNN (MTL) \cite{neumann2019improving} & 60.15& - \\\hline
CNN (MTL) \cite{latif2020multi}& 65.6$\pm$2.0 &59.5$\pm$2.4\\\hline
Semi-supervised AAE (MTL) \cite{latif2020multi}    &66.7$\pm$1.4     &  60.3$\pm$1.1           \\ \hline
CNN-BLSTM$_{\scriptsize{(\text{baseline})}}$ (STL)&  64.3$\pm$1.9       &   57.2$\pm$2.1          \\ \hline
\begin{tabular}[c]{@{}l@{}}CNN-BLSTM$_{\scriptsize{(\text{baseline})}}$ \\(STL\scriptsize{+augmentations})\end{tabular}&  65.1$\pm$1.8       &   58.5$\pm$1.7          \\ \hline
MTL-AUG & \textbf{68.1$\pm$1.5}        &\textbf{ 61.4$\pm$ 0.9 }           \\ \hline
MTL-AUG (additional data) & \textbf{68.7$\pm$1.3}        &\textbf{ 62.1$\pm$ 1.2 }           \\ \hline
\end{tabular}
\label{table:within}
\end{table}



\subsection{Cross-Corpus and Cross-Language Evaluations}
\label{cross-corpus}
\subsubsection{Cross-Corpus:}In this experiment, we perform a cross-corpus analysis to verify the generalisability of the proposed framework. We trained models on IEMOCAP, and testing is performed on the MSP-IMPROV data. We choose IEMOCAP as training data, since it is more balanced than other corpora. The other reason to select this scheme is for comparison with existing studies, which decided for a similar 
training~\cite{latif2020multi,neumann2019improving,sahu2018enhancing}. We select 30\,\% of the MSP-IMPROV data for parameter selection and 70\,\% as testing data. The training and testing data are randomly selected.

\begin{table}[!ht]
\centering
\caption{Cross-corpus evaluation results for emotion recognition.}
\begin{tabular}{|l|l|}
\hline
Model                                         & UAR (\%)\\ \hline
\begin{tabular}[c]{@{}l@{}}Attentive CNN (MTL) \cite{neumann2019improving}\end{tabular} &45.7\\ \hline
\begin{tabular}[c]{@{}l@{}}Conditional-GAN (STL) \cite{sahu2018enhancing}\end{tabular} &45.4 \\ \hline
\begin{tabular}[c]{@{}l@{}}Semi-supervised AAE (MTL) \cite{latif2020multi} \end{tabular} & 46.4$\pm$0.32 \\ \hline
\begin{tabular}[c]{@{}l@{}}CNN-BLSTM (STL)$_{\scriptsize{(\text{baseline})}}$ \end{tabular} & 45.4$\pm$0.83  \\ \hline
\begin{tabular}[c]{@{}l@{}}CNN-BLSTM (STL)$_{\scriptsize{(\text{baseline})}}$ (\scriptsize{+ augmentations}) \end{tabular} & 46.2$\pm$1.3 \\ \hline
\begin{tabular}[c]{@{}l@{}}MTL-AUG \end{tabular} & \textbf{47.2$\pm$0.41 } \\ \hline
\begin{tabular}[c]{@{}l@{}}MTL-AUG (additional data) \end{tabular} & \textbf{48.1$\pm$0.30 }\\ \hline
\end{tabular}
\label{cross}
\end{table}

We compare our results with different studies in Table \ref{cross}. In \cite{neumann2019improving}, the authors utilise the representations learnt from unlabelled data and feed it to an attention-based multitask CNN classifier. They show that the classifier's performance can be improved by using the representations from unlabelled data. In \cite{sahu2018enhancing}, the authors use the synthetic data generated by a generative adversarial network (GAN)
to augment the emotional classifier. They show that augmentation can improve the generalisation that leads to performance improvement. A recent study \cite{latif2020multi} utilised a semi-supervised AAE in an MTL setting to improve the generalisation of SER systems. 
They use supervised auxiliary tasks, including speaker and gender identification. The authors show that the generalisation of SER systems can be improved by learning the speaker and gender information from the data. In contrast, our proposed MTL-AUG framework learns the generalised representations from the augmented data by learning augmentation-types changes applied to the data. These generalised representations help achieve improved results for cross-corpus SER. 


\subsubsection{Cross-Language:} We also evaluate the MTL-AUG setup on cross-language SER. For this experiment -- as outlined above -- we use the IEMOCAP and EMODB corpora. We compare the results with \cite{kim2017towards} for cross-language SER, where the authors used a multitask LSTM model with gender and naturalness as auxiliary tasks. The results of the comparison are presented in Table \ref{tab:cross_lang}. We train the models on IEMOCAP (English), and EMODB (German) is used for validation and testing for four class emotion classification. Similar to the cross-corpus experiments, we also achieve improved results for cross-language SER.

\begin{table}[!ht]
\centering
\scriptsize
\caption{Cross-language evaluation results (UAR \%) for emotion recognition.}
\begin{tabular}{|l|l|l|}
\hline
Model & \begin{tabular}[c]{@{}l@{}}IEMOCAP \tiny{(English)}\\ to EMODB \tiny{(German)}\end{tabular} & \begin{tabular}[c]{@{}l@{}}EMODB \tiny{(German)}\\ to IEMOCAP \tiny{(English)}\end{tabular} \\ \hline
MTL-LSTM \cite{kim2017towards}&   43.4$\pm$1.8    &39.1$\pm$1.6 \\ \hline
CNN-BLSTM (STL)\tiny{(baseline)}   &   42.1$\pm$ 1.9                    &  38.4$\pm$ 1.8                                                                            \\ \hline
\begin{tabular}[c]{@{}l@{}}CNN-BLSTM (STL)\tiny{(baseline)}\\ \tiny{(+ augmentations)}\end{tabular}& 43.6$\pm$1.5            &     39.5$\pm$ 1.7 \\\hline

MTL-AUG   &   \textbf{45.7$\pm$1.3}       &      \textbf{42.1$\pm$1.6 }          \\ \hline
MTL-AUG \tiny{(additional data)}   & \textbf{46.8$\pm$1.4 }                &       \textbf{41.5$\pm$  1.6     }                                                                 \\ \hline

\end{tabular}
\label{tab:cross_lang}
\end{table}


\subsection{Evaluation of robustness to noise}
\label{sec:noisy}
In this experiment, we evaluate the proposed model in noisy conditions. We compare our results with a recent study \cite{latif2020deep} that applies a deep architecture to learn a robust representation and exploits a combination of mixup and speed perturbation data augmentation techniques to achieve improved generalisation. We consider the same settings chosen in \cite{latif2020deep} and train the model on clean data and evaluate on noisy samples. For a fair comparison with \cite{latif2020deep}, we select the same three signal-to-noise ratio (SNR) values [0, 10, 20] and select five noises, including kitchen, park, cafeteria, station, and traffic, from the DEMAND dataset. These noises are randomly added to the testing data at three SNR values [0, 10, 20]. We also implemented models used \cite{huang2017deep,wijayasingha2021robustness} for robust SER to extend our comparison scope. In \cite{huang2017deep}, authors use attentive CNN-BLSTM model for robust SER. Similarly, authors in \cite{wijayasingha2021robustness} use attention based CNN model to perform noise robust SER. Results on the IEMOCAP data are compared with \cite{latif2020deep,huang2017deep,wijayasingha2021robustness} and the baseline in Table \ref{noisy}. 

\begin{table}[!ht]
\centering
\caption{Comparing the proposed model against noisy condition with state-of-the-art architectures.}
\begin{tabular}{|l|ccl|}
\hline
\multirow{2}{*}{Model} & \multicolumn{3}{c|}{UAR (\%)}                         \\ \cline{2-4} 
                       & \multicolumn{1}{c|}{0 dB} & \multicolumn{1}{c|}{10} & 20 \\ \hline
\begin{tabular}[c]{@{}l@{}}DenseNet (STL) \\(+augmentations) \cite{latif2020deep} \end{tabular}&\multicolumn{1}{c|}{34.2$\pm$ 1.2}  & \multicolumn{1}{c|}{40.9$\pm$1.5}   &43.1 $\pm$1.1    \\ \hline

\begin{tabular}[c]{@{}l@{}}CNN-BLSTM +attention (STL)\\ \cite{huang2017deep}\end{tabular}   & \multicolumn{1}{c|}{34.0$\pm$1.5}  & \multicolumn{1}{c|}{40.1$\pm$1.6}   & 41.8 $\pm$1.7     \\ \hline

\begin{tabular}[c]{@{}l@{}}CNN +attention (STL) \cite{wijayasingha2021robustness}\end{tabular}   & \multicolumn{1}{c|}{33.4$\pm$1.8}  & \multicolumn{1}{c|}{39.8$\pm$2.1}   & 41.9 $\pm$1.7     \\ \hline

CNN-BLSTM (STL) $_{\scriptsize{(\text{baseline})}}$   & \multicolumn{1}{c|}{33.5$\pm$1.5}  & \multicolumn{1}{c|}{39.2$\pm$1.4}   & 41.7 $\pm$1.4     \\ \hline
\begin{tabular}[c]{@{}l@{}}CNN-BLSTM (STL)$_{\scriptsize{(\text{baseline})}}$\\ (\scriptsize{+ augmentations}) \end{tabular}   & \multicolumn{1}{c|}{34.8$\pm$1.3}  & \multicolumn{1}{c|}{41.2$\pm$1.5}   & 43.8 $\pm$1.6     \\ \hline

MTL-AUG & \multicolumn{1}{c|}{\textbf{37.5$\pm$1.0}}  & \multicolumn{1}{c|}{\textbf{43.2$\pm$1.3}}   &\textbf{45.1$\pm$1.3}    \\ \hline
MTL-AUG (additional data) & \multicolumn{1}{c|}{\textbf{39.1$\pm$1.3}}  & \multicolumn{1}{c|}{\textbf{44.1$\pm$1.4}}   & \textbf{46.5$\pm$1.3}   \\ \hline
\end{tabular}
\label{noisy}
\end{table}

In contrast to the deep networks used in  \cite{latif2020deep,huang2017deep,wijayasingha2021robustness} and baseline, we achieve better results. This shows that the proposed MTL approach enables the MTL-AUG to learn generalised representations, which help achieve robustness to perform SER in noisy conditions. Both ``baseline (+augmentation)'' and the deep DenseNet used in \cite{latif2020deep} are trained in STL setting exploiting the augmented data. We show in Table \ref{adv} that training the STL model with augmented data helps improve robustness against noisy conditions; however, these models do not have access to the latent information available in the augmented data. We use this extra information in our proposed MTL-AUG model, where we perform augmentation-type classification as an auxiliary task to exploit the augmented data in the MTL setting.

\subsection{Adversarial Attacks}
\label{adver}
In adversarial settings, we choose two adversarial attacks, including the Fast Gradient Sign Method (FGSM) \cite{GoodfellowSS14} and the Basic Iterative Method (BIM) \cite{KurakinGB17a} to evaluate the robustness of MTL-AUG. FGSM generates adversarial samples by adding a scaled perturbation in the direction of the gradient of the loss function. The BIM attack builds upon the FGSM attack by applying it multiple times iteratively with small $\epsilon$ instead of applying the adversarial noise in a single step. We apply these two attacks with the perturbation factor $\epsilon$ = $0.08$, and the performance is reported in Table \ref{adv}. We compare our results with that of \cite{latif2020deep}, where the authors consider the same adversarial attacks. In addition, we also use the implementation of robust models use in  \cite{huang2017deep,wijayasingha2021robustness} for evaluation against adversarial attacks. Comparisons show that we achieve better performance than these existing studies.
\begin{table}[!ht]
\centering
\caption{Performance (UAR \%) comparison against adversarial attacks using different models. MTL-AUG
(additional data) represents when additional unlabelled data is used. }
\begin{tabular}{|l|cc|}
\hline
\multirow{2}{*}{Model} & \multicolumn{2}{c|}{Adversarial Attacks} \\ \cline{2-3} 
  & \multicolumn{1}{c|}{FSGM}      & BIM     \\ \hline
DenseNet (STL) (+augmentations) \cite{latif2020deep} & \multicolumn{1}{c|}{44.0$\pm$ 1.1}          &  36.4$\pm$1.3       \\ \hline
CNN-BLSTM +attention (STL) \cite{huang2017deep} & \multicolumn{1}{c|}{43.8$\pm$ 1.5}          &  37.2$\pm$1.5      \\ \hline
CNN +attention (STL) \cite{wijayasingha2021robustness}& \multicolumn{1}{c|}{42.7$\pm$ 1.7}          &  36.7$\pm$1.4      \\ \hline
CNN-BLSTM (STL) $_{\scriptsize{(\text{baseline})}}$  & \multicolumn{1}{c|}{42.5$\pm$1.5}          &  35.8$\pm$ 1.6       \\ \hline
\begin{tabular}[c]{@{}l@{}}CNN-BLSTM (STL) $_{\scriptsize{(\text{baseline})}}$ \\(\scriptsize{+ augmentations}) \end{tabular}   & \multicolumn{1}{c|}{44.6$\pm$1.4}          &  37.8$\pm$1.4       \\ \hline
MTL-AUG  & \multicolumn{1}{c|}{\textbf{46.2$\pm$1.2}}          & \textbf{39.1$\pm$1.4 }       \\ \hline
MTL-AUG (additional data)  & \multicolumn{1}{c|}{\textbf{47.5$\pm$1.0} }         &\textbf{40.6$\pm$1.2 }        \\ \hline
\end{tabular}
\label{adv}
\end{table}

 In \cite{latif2020deep}, the authors develop a deep architecture to learn a robust representation. In addition, they utilise speed perturbation and mixup augmentation in the STL setting to achieve generalisation. In contrast, we select augmentation-type classification as an auxiliary task in the MTL scenario. This facilitates generalisation in the network by learning the common representations for both primary and auxiliary tasks. 

\subsection{Selection of Data Augmentation}
\label{augment}
\begin{figure*}[!ht]%
\centering
\begin{subfigure}{0.45\linewidth}
\includegraphics[trim=0cm 0cm 0cm 0cm,clip=true,width=\linewidth]{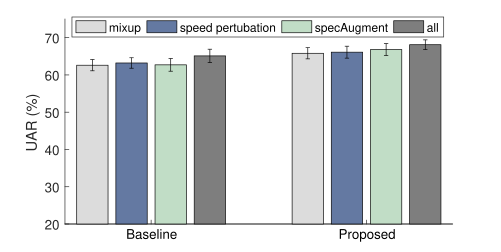}%
\captionsetup{justification=centering}
\caption{IEMOCAP}%
\label{IEMOCAPAUG}%
\end{subfigure}\hfill%
\begin{subfigure}{0.45\linewidth}
\includegraphics[trim=0.0cm 0cm 0cm 0cm,clip=true,width=\linewidth]{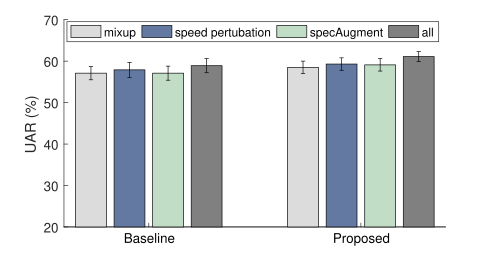}%
\captionsetup{justification=centering}
\caption{MSP-IMPROV} %
\label{MSPAUG}%
\end{subfigure}\hfill
\caption{Results using single augmentation in the auxiliary task of augmentation-type  classification vs all.}
\label{fig:aug}
\end{figure*}

In this experiment, we evaluate the model using different schemes in the auxiliary task of augmentation-type classification. We start with single augmentation and perform binary classification (augmented or not augmented) in the auxiliary task using different data augmentation techniques. Results are plotted in Figure \ref{fig:aug}, which highlight that the performance of the MTL model with a single augmentation-type in the augmentation-type classifier is poorer than using multiple augmentation-types classification. This shows that giving the model more diverse augmented data helps to learn generalised representations compared to learning to classify single data augmentation. 

\subsection{Size of Labelled Data}
\label{laballed}
\begin{figure}[!ht]
\centering
\includegraphics[width=0.47\textwidth]{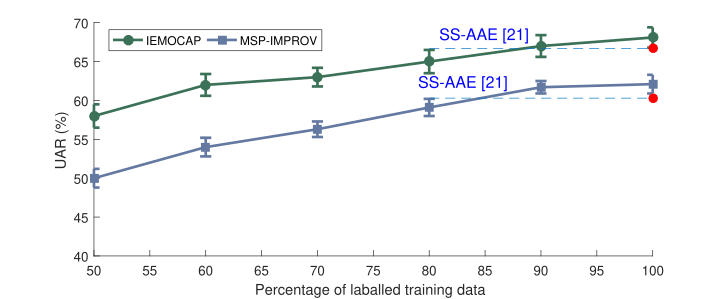}
\caption{Results for SER (UAR \%) with changing the amount of labelled training data in IEMOCAP and MSP-IMPROV.  }
\label{fig:data}
\end{figure}

In this experiment, we change the amount of labelled data for training the models, and the results are compared with a semi-supervised AAE (SS-AAE) \cite{latif2020multi}. We present the outcomes on IEMOCAP and MSP-IMPROV in Figure \ref{fig:data}. We plot the results with different percentages of labelled training data. The proposed framework improves the SER performance considerably against the baseline CNN-BLSTM.  
We also compare the results with SS-AAE \cite{latif2020multi} on the SER performance. Results are plotted in Figure \ref{fig:data}, where the red dot shows the performance achieved by SS-AAE \cite{latif2020multi} using 100\,\% of source data along with the unlabelled data of  LibriSpeech. We achieve similar performance using 80-86\,\% of labelled training data as highlighted by a dotted blue line. This shows that the proposed MTL-AUG effectively learns the emotional representation from augmented data to improve the performance while reducing the required labelled data. 

\subsection{Ablation Experiments}\cite{shen2011non}
In this experiment, we validate the necessity and effectiveness of each module integrated with our proposed framework. Results are presented in Table 7. This experiment starts with the proposed framework containing all components, including the attention layer, centre loss, auxiliary augmentation-type classifier, and reconstruction decoder. We remove the auxiliary augmentation-type classifier and reconstruction decoder in models 2 and 3. We keep removing different components until we obtain a simple CNN-BLSTM (model 5) classifier without the attention, centre loss, augmentation-type classifier, and reconstruction decoder. We use model 4 as baseline classifier in other sections \ref{within}-\ref{laballed}. There is a considerable drop in UAR (\%) when one or more modules are removed from the proposed framework. When an STL CNN-BLSTM classifier (module 5) is used, we see a considerable performance drop for both within and cross-corpus SER. This shows that the STL CNN-BLSTM cannot learn better generalisation compared to the MTL framework using the augmentation-type classifier, the reconstruction decoder, or both as auxiliary tasks. This shows that auxiliary tasks promote generalised representations in the network by learning the shared representations. Overall, these ablation experiments show that all the proposed model components are chosen carefully to improve the SER performance effectively.  

\begin{table*}[!ht]
\centering
\tiny
\caption{Results (UAR \%) for within-corpus and cross-corpus settings using different configurations of the proposed model.}
\begin{tabular}{|l|c|ll|l|l|ll|l|}
\hline
\multirow{2}{*}{Model} &Configuration & \multicolumn{2}{c|}{Auxiliary tasks}                                                                     & \multirow{2}{*}{Centre loss} & \multirow{2}{*}{Attention} & \multicolumn{2}{c|}{\begin{tabular}[c]{@{}c@{}}Within corpus\\         UAR(\%)\end{tabular}} & \multicolumn{1}{c|}{\multirow{2}{*}{\begin{tabular}[c]{@{}c@{}}Cross-corpus\\        UAR(\%)\end{tabular}}} \\ \cline{3-4} \cline{7-8}
                      && \multicolumn{1}{c|}{\begin{tabular}[c]{@{}c@{}}Augmentation- \\ type classifier\end{tabular}} & Reconstruction &                              &                            & \multicolumn{1}{l|}{IEMOCAP}                          & MSP-IMPROV   & \multicolumn{1}{c|}{}                                                                                  \\ \hline
                       
1  &\begin{minipage}{.4\textwidth}
      \includegraphics[width=0.9\linewidth,height=50mm]{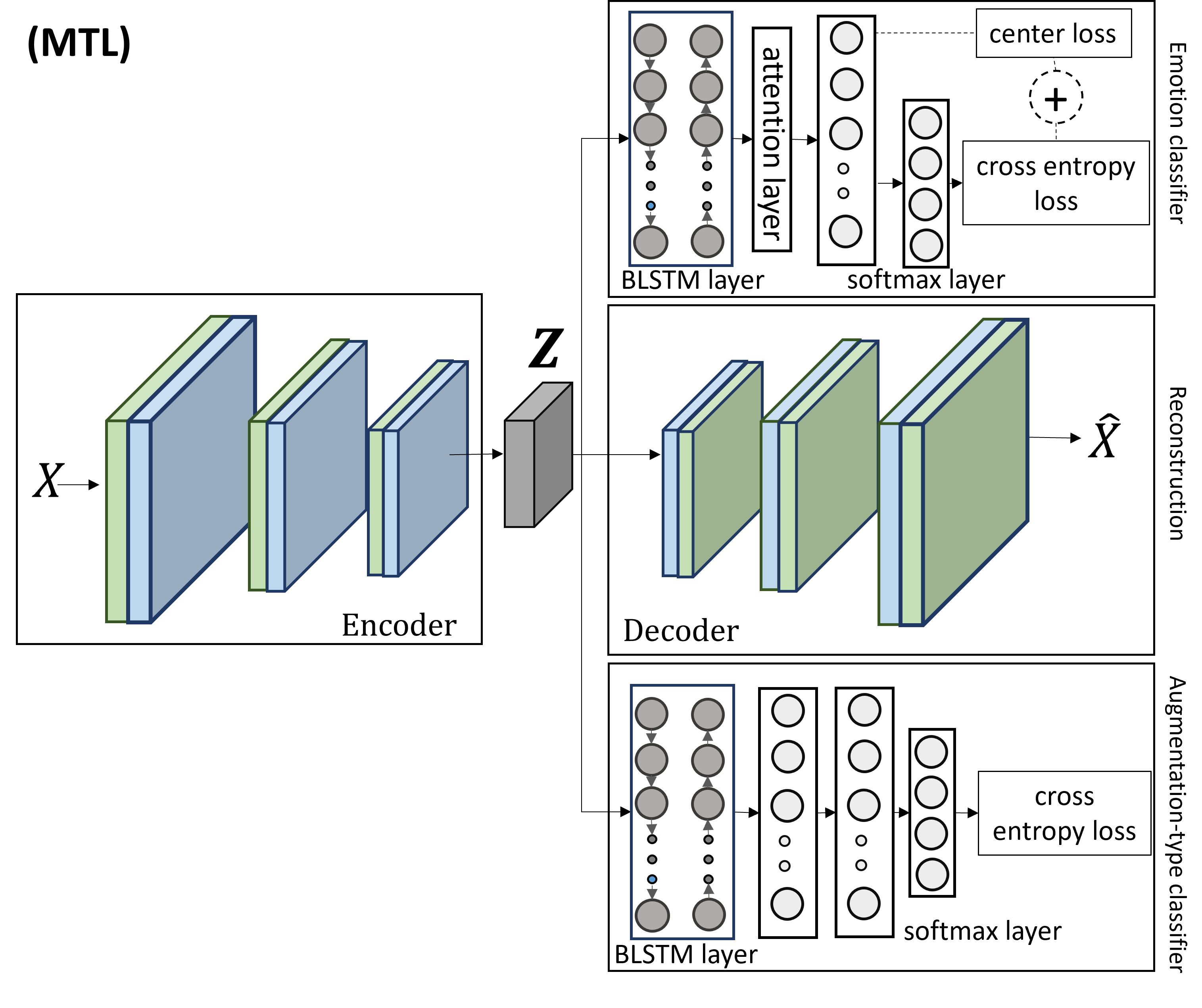}\end{minipage}& \multicolumn{1}{l|}{\checkmark} &   \checkmark  &  \checkmark &   \checkmark  & \multicolumn{1}{l|}{\textbf{68.1$\pm$1.5}}  &   \textbf{62.1$\pm$1.2}   &  \textbf{47.2$\pm$0.41 }                              \\ \hline
2  &\begin{minipage}{.4\textwidth}
      \includegraphics[width=0.85\linewidth,height=35mm]{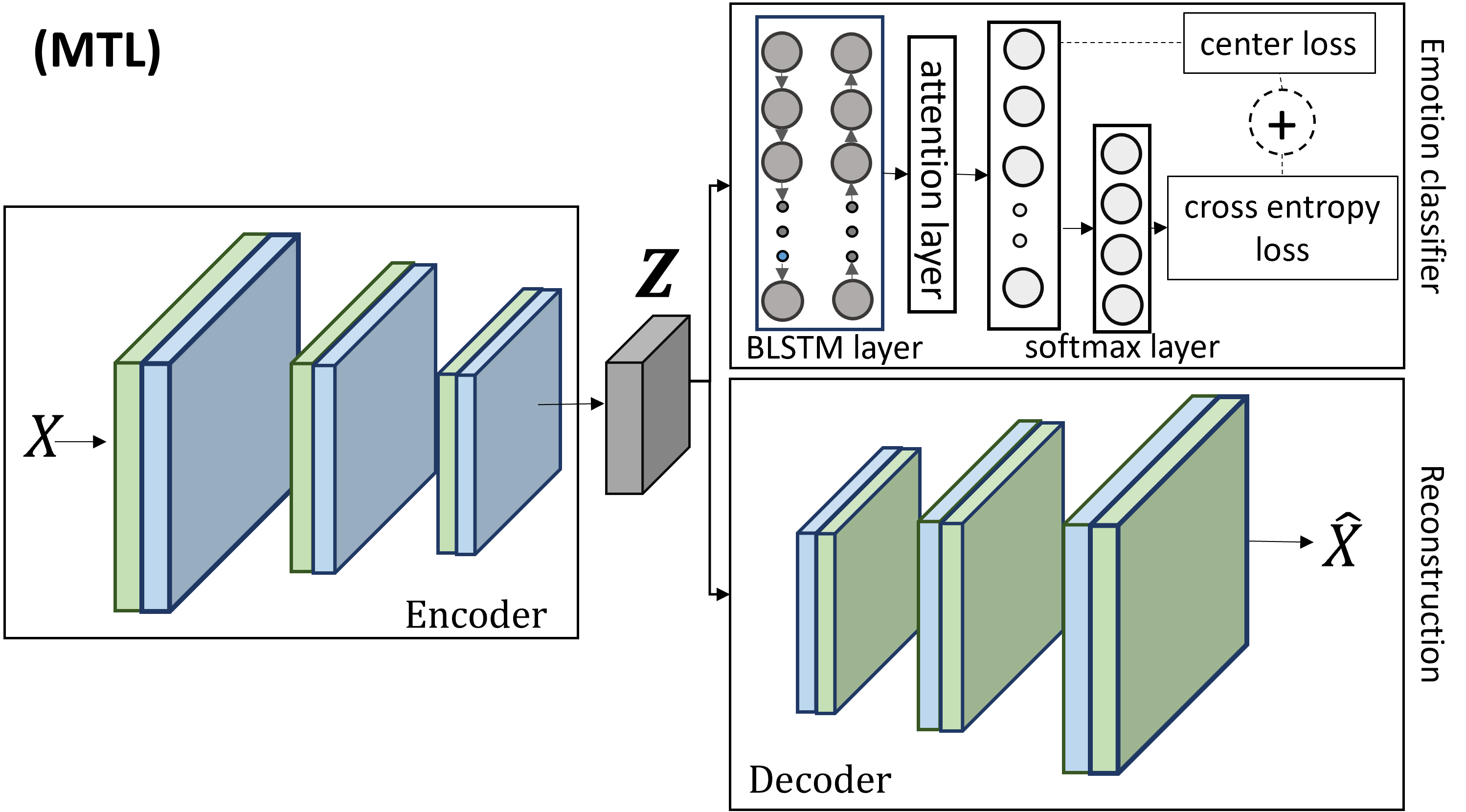}\end{minipage}& \multicolumn{1}{l|}{\xmark} &   \checkmark  &  \checkmark   &   \checkmark   & \multicolumn{1}{l|}{66.7$\pm$1.5} &   60.5$\pm$1.4      & 46.2$\pm$0.81       \\ \hline     
3  &\begin{minipage}{.4\textwidth}
      \includegraphics[width=0.8\linewidth,height=18mm]{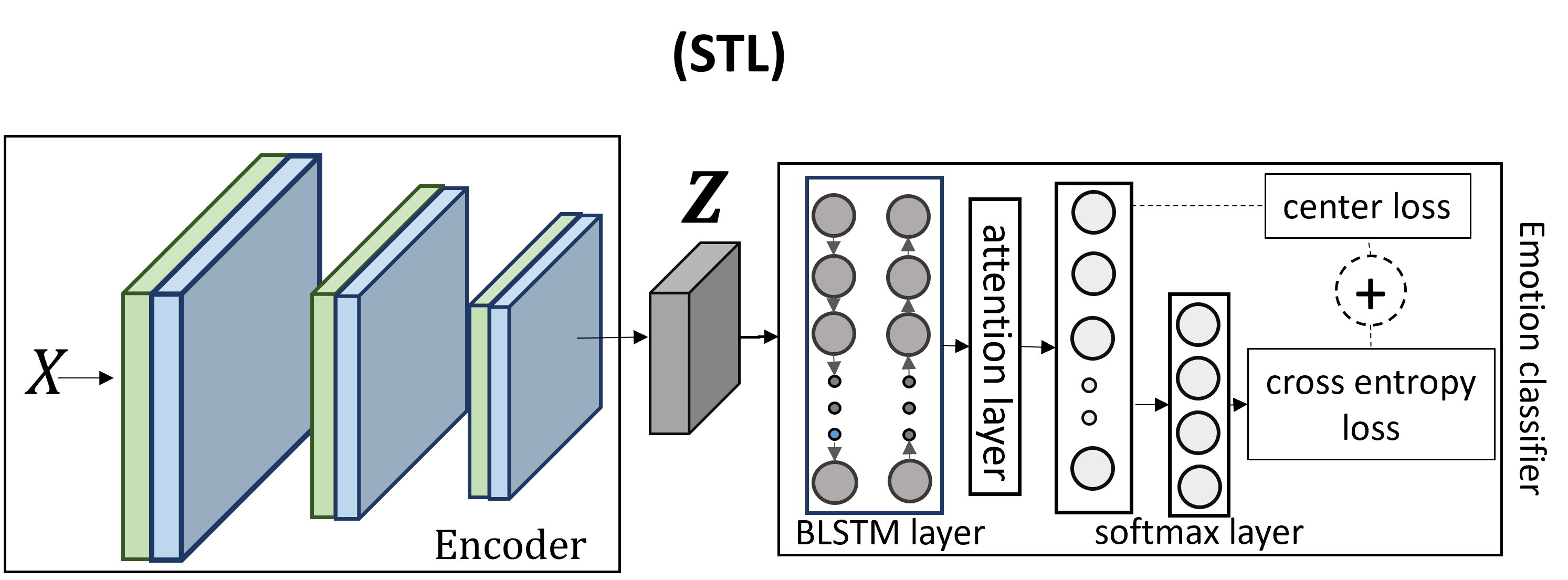}\end{minipage}& \multicolumn{1}{l|}{\xmark} &  \xmark  &   \checkmark  &  \checkmark   & \multicolumn{1}{l|}{65.1$\pm$1.7} &   59.0$\pm$1.8        & 45.8$\pm$1.0     \\ \hline  
4   &\begin{minipage}{.4\textwidth}
      \includegraphics[width=0.8\linewidth,height=18mm]{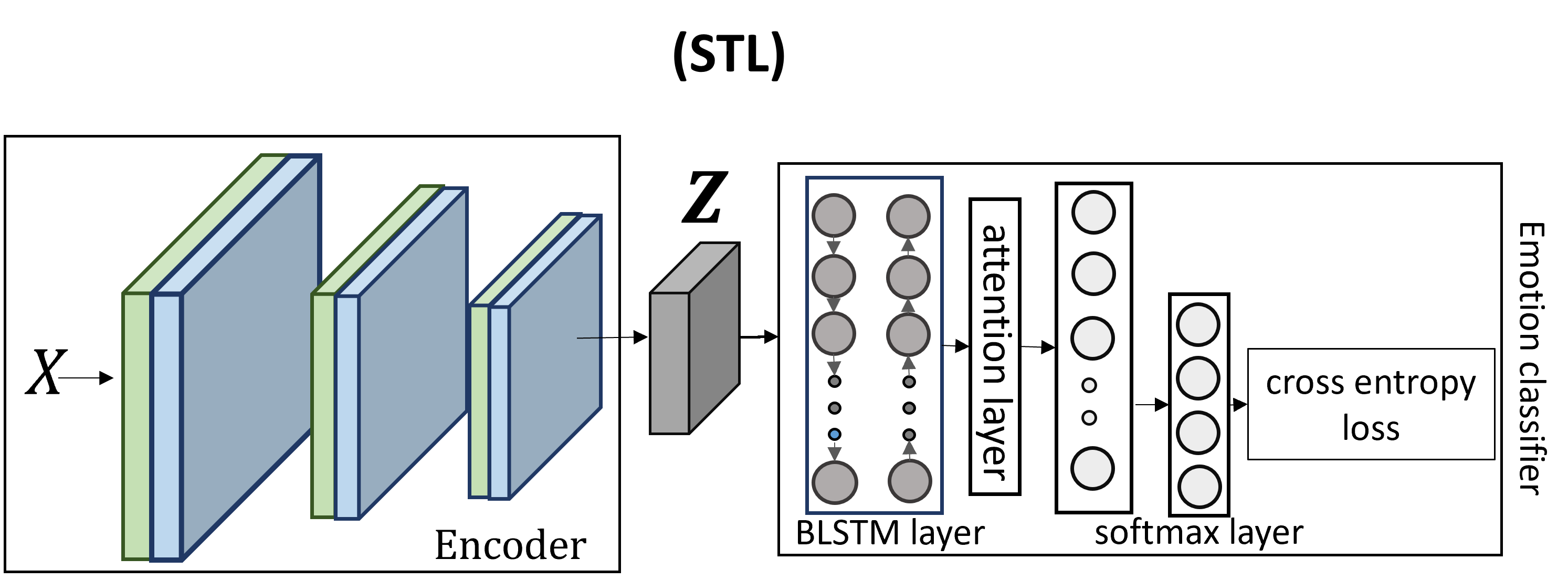}\end{minipage}& \multicolumn{1}{l|}{\xmark}   &  \xmark   &   \xmark &\checkmark   & \multicolumn{1}{l|}{64.3$\pm$1.9} &   58.2$\pm$2.1        &  45.4$\pm$1.2         \\ \hline  
5  &\begin{minipage}{.4\textwidth}
      \includegraphics[width=0.8\linewidth,height=18mm]{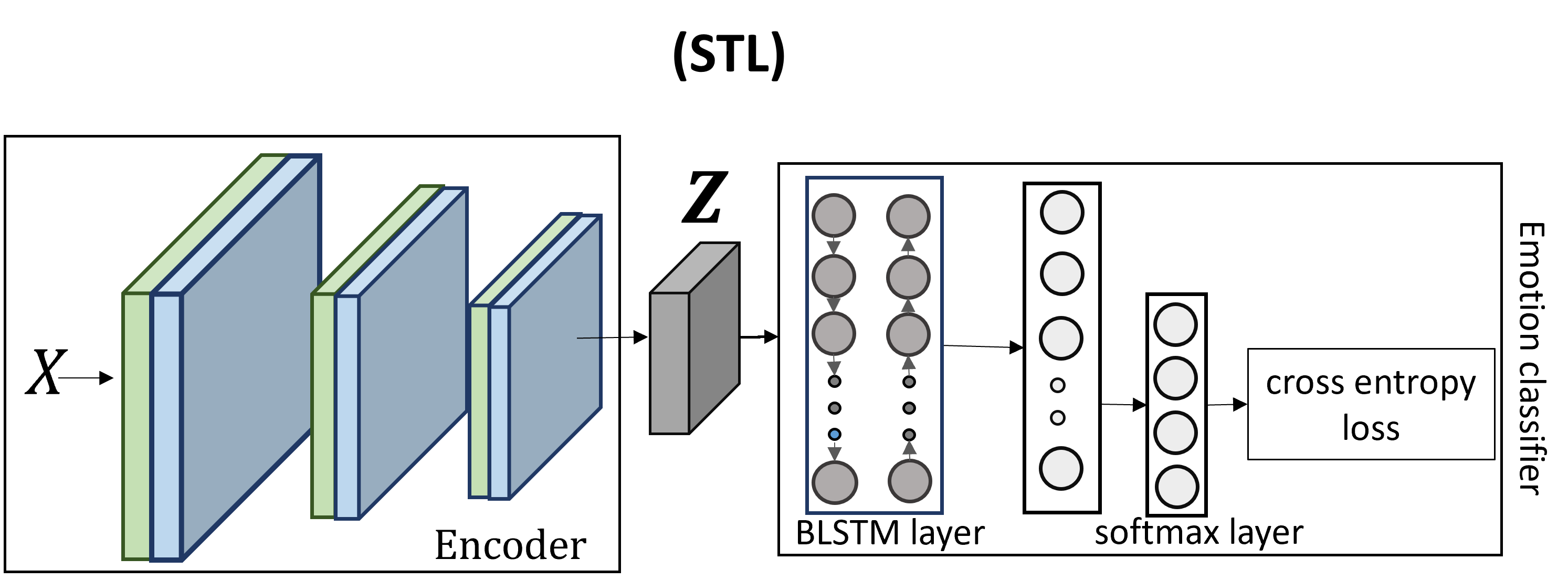}\end{minipage}& \multicolumn{1}{l|}{\xmark} &  \xmark  &  \xmark &  \xmark   & \multicolumn{1}{l|}{62.8$\pm$2.1}   &   56.5$\pm$1.9     &     43.6$\pm$1.5    \\ \hline

\end{tabular}
\end{table*}

\section{Conclusions and Outlook}
This contribution addressed the open challenge of improving the generalisation of speech emotion recognition (SER) with novel auxiliary tasks that do not require any additional labels for training a multi-task learning (MTL) model. We proposed augmentation-type classification and reconstruction as auxiliary tasks that minimise the required labelled data by effectively utilising the information available in the augmented data and facilitating the utilisation of unlabelled data in a semi-supervised way. 
The key highlights are as follows:
\begin{itemize}
    \item The multi-task model offers improved within-corpus, cross-corpus, and cross-language
    emotion classification. It also shows improved generalisation against noisy speech and adversarial attacks. This is due to the proposed auxiliary tasks that helps the model learn shared representations from augmented data. 
    \item Considerable improvements in results were found when additional unlabelled data was incorporated into the proposed MTL semi-supervised framework. This helped the model to regulate the generalised representations by learning from unlabelled data. 
    \item We were able to reduce the amount of labelled training data by more than  10\,\% while achieving a similar performance reported by a recent related study \cite{latif2020multi} using 100\,\% training data.
\end{itemize}
Future work includes exploring multi-model auxiliary tasks to improve the primary task of speech emotion recognition by learning generalised representation.

\end{document}